\documentstyle[emulateapj,psfig,natbib,amsfonts,amsmath]{article}

\def\nod{\nodata}
\def\grb{GRB\,030226}

\def\prince{1}
\def\ociw{2}
\def\hubble{3}
\def\pom{4}
\def\psu{5}
\def\uh{6}
\def\cit{7}
\def\uhh{8}
\def\gemini{9}
\def\hia{10}
\def\eso{11}

\begin{document}

\title{High-Resolution Spectroscopy of GRB\,030226:  Features of a
Massive Star Progenitor or Intervening Absorption Systems?}

\author{
Min-Su~Shin\altaffilmark{\prince},
Edo~Berger\altaffilmark{\ociw,}\altaffilmark{\prince,}\altaffilmark{\hubble},
Bryan~E.~Penprase\altaffilmark{\pom},
Derek~B.~Fox\altaffilmark{\psu},
Paul~A.~Price\altaffilmark{\uh},
Shri~R.~Kulkarni\altaffilmark{\cit},
Alicia~M.~Soderberg\altaffilmark{\cit},
Michael~J.~West\altaffilmark{\uhh,}\altaffilmark{\gemini},
Patrick~C\^ot\'e\altaffilmark{\hia},
Andr\'es~Jord\'an\altaffilmark{\eso}
}

\altaffiltext{\prince}{Princeton University Observatory,
Peyton Hall, Ivy Lane, Princeton, NJ 08544}

\altaffiltext{\ociw}{Observatories of the Carnegie Institution
of Washington, 813 Santa Barbara Street, Pasadena, CA 91101}

\altaffiltext{\hubble}{Hubble Fellow}

\altaffiltext{\pom}{Pomona College Department of Physics and Astronomy,
610 N. College Avenue, Claremont, CA 91711}

\altaffiltext{\psu}{Department of Astronomy and Astrophysics,
Pennsylvania State University, 525 Davey Laboratory, University
Park, PA 16802}

\altaffiltext{\uh}{Institute for Astronomy, University of Hawaii,
2680 Woodlawn Drive, Honolulu, HI 96822}

\altaffiltext{\cit}{Division of Physics, Mathematics and Astronomy,
105-24, California Institute of Technology, Pasadena, CA 91125}

\altaffiltext{\uhh}{Department of Physics and Astronomy, University 
of Hawaii, Hilo, HI 96720}

\altaffiltext{\gemini}{Gemini Observatory, Casilla 603, La Serena, Chile}

\altaffiltext{\hia}{Herzberg Institute of Astrophysics, National Research Council of Canada, 
5071 West Saanich Road, Victoria, BC V8X 4M6, Canada}

\altaffiltext{\eso}{European Southern Observatory, Karl-Schwarzschild-Strasse 
2, 85748 Garching, Germany}

\begin{abstract} 
We present a high-resolution Keck/ESI spectrum of \grb, which exhibits
four absorption systems at $z=1.04329$, $1.95260$, $1.96337$, and
$1.98691$.  The two highest redshift systems, separated by about 2400
km s$^{-1}$, have been previously suspected as kinematic features
arising in the circumstellar wind around the progenitor star.
However, the high column densities of low-ionization species
(including possibly neutral hydrogen) in the blue-shifted system, are
inconsistent with the expected highly ionized state of the
circumstellar wind from the massive progenitor star, even prior to the
GRB explosion.  This conclusion is also supported by the lack of
detectable absorption from fine-structure transitions of \ion{Si}{2}
and \ion{Fe}{2}.  Instead we conclude that the two redshift systems
are similar to multiple DLAs found in QSO sight lines with a similar
velocity separation and chemical abundance of [Cr/Fe] and [Zn/Fe].
The absorption system at $z=1.96337$ is likely an intervening low-mass
galaxy, possibly related to the GRB host as part of a forming
large-scale structure.
\end{abstract}


\section{Introduction}
\label{sec:intro}

The idea that the progenitors of long-duration $\gamma$-ray bursts
(GRBs) are massive stars is now well-established, based in particular
on the detection of associated type Ic supernovae (e.g.,
\citealt{gvv+98,smg+03}).  The nature of these supernovae, and
theoretical considerations, have lead to the suggestion that the
progenitors are massive Wolf-Rayet (WR) stars \citep{mw99}.  Such
stars shed a considerable fraction of their mass in a fast ($\sim
10^3$ km s$^{-1}$) wind (e.g., \citealt{abb78}).  As a result,
absorption spectroscopy of GRB afterglows can in principle probe the
composition and kinematics of the wind, providing in turn direct
information on the nature and identity of the progenitor star.
Previous observations have in fact uncovered such blue-shifted
absorption systems, which have been used to argue for carbon-rich (WC)
Wolf-Rayet progenitors (e.g., \citealt{mhc+03,fdl+05,bpc+06}),
potentially in binary systems \citep{swh+05,bpc+06}.

One central question to answer in applying the potentially powerful
diagnostics from absorption lines is whether blue-shifted absorbers
arise from outflowing circumstellar gas, or separate intervening
absorbers arising from different galaxies.  In addition, the intense
radiation field produced by the GRB can modify the circumstellar
environment through ionization, excitation, and destruction of dust
grains (e.g., \citealt{pl98,dh02}).  However, these processes, as well
as the relative abundance of low- and high-ionization ions, may in
fact serve to distinguish the location of the absorbers \citep{pcb06}.

Here we analyze a high resolution spectrum of \grb\ in a continued
effort to characterize the circumstellar and interstellar environments
around GRBs.  This burst is of particular interest because it is one
of only a few that exhibit a blue-shifted absorption system with a
velocity of a ${\rm few}\times 10^3$ km s$^{-1}$ \citep{kgr+04},
possibly representing the circumstellar wind from the progenitor star.
However, based on the relative strengths of the low- and
high-ionization lines, and the lack of detectable fine-structure
transitions we conclude that the blue-shifted absorption system is not
circumstellar.  Instead we conclude that the blue-shifted system is an
intervening galaxy similar to the multiple DLAs found in some quasar
spectra \citep{el01,le03}.

\section{Observations}
\label{sec:obs}

\grb\ was detected by the HETE-2 satellite on 2003 February 26.157 UT.
The optical afterglow was discovered with the du Pont 100-inch
telescope at Las Campanas Observatory 2.6 hr after the burst
\citep{gcn1879}, followed by a measurement of the redshift, $z=1.986$
\citep{kgr+04}.  The latter authors present an analysis of a low
resolution spectrum obtained with the VLT, which also exhibits a
second absorption system at $z=1.962$.

We obtained a 2700 s spectrum of the afterglow with the Echellette
Spectrograph and Imager (ESI) mounted on the Keck II 10-m telescope
starting 4.2 hr after the burst.  The data were reduced using custom
IRAF routines to bias-subtract, flat-field, and rectify the ten
individual echelle orders.  Sky subtraction was performed using the
method and software described in \citet{kel03}.  Wavelength
calibration was performed using CuAr and HgNeXe arc lamps and
air-to-vacuum and heliocentric corrections were applied.  The spectrum
covers the range of $3900-10500$ \AA\ at a velocity resolution of 11.5
km s$^{-1}$.  The signal-to-noise over the bulk of the spectral range
($4500-9500$ \AA) is about 40.

We obtained a second spectrum, totaling 6000 s, to search for
variability in the wavelengths and equivalent widths with ESI 27.4 hr
after the burst.  The data were reduced in the manner detailed above.
The signal-to-noise of this second spectrum is about 10.

\begin{figure}
\epsscale{1}
\plotone{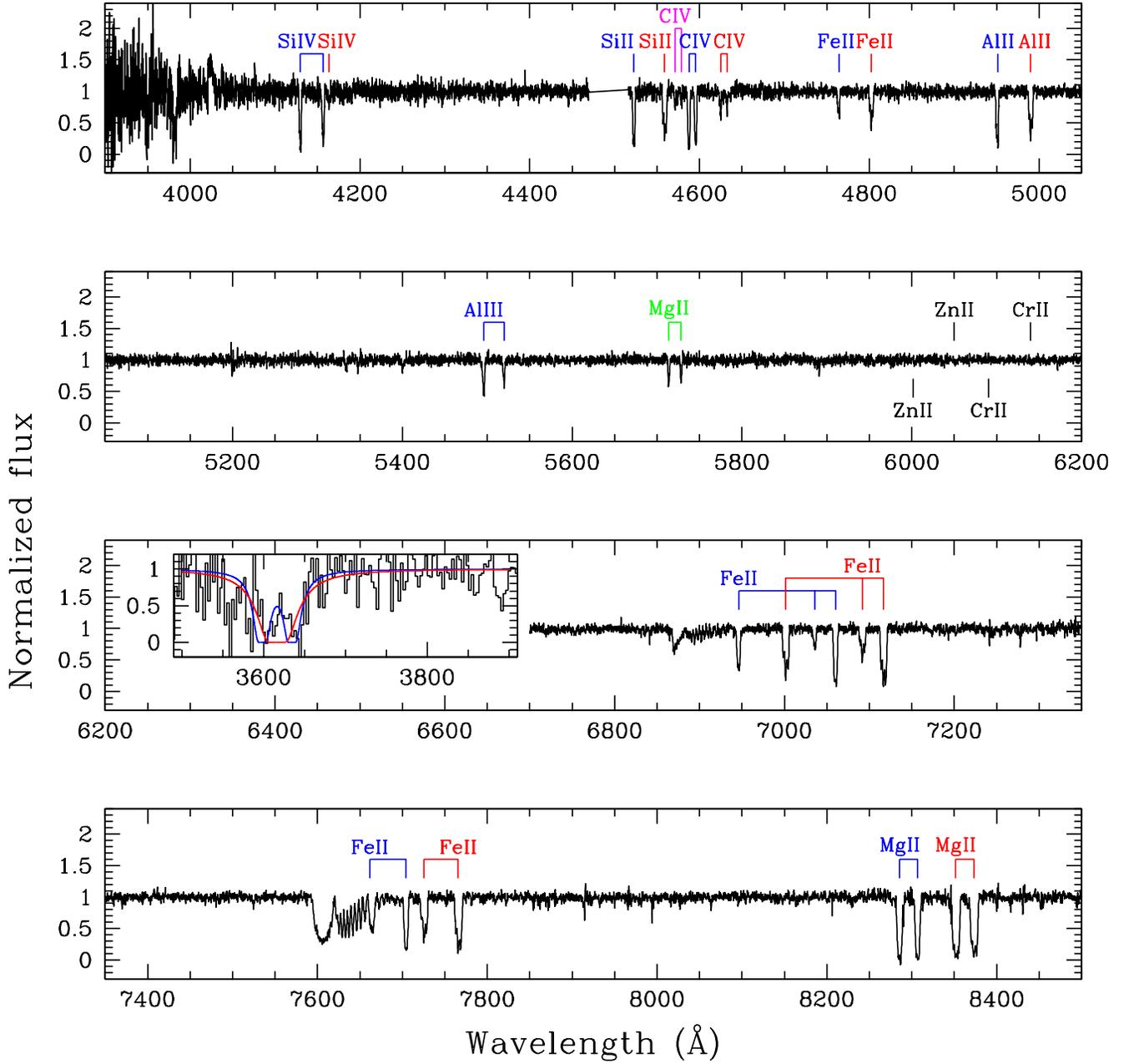}
\caption{Normalized spectrum of GRB 030226 afterglow obtained on 2003
Feb.~26.333 UT (4.22 hours after the burst) and ({\it inset}) the part
of the VLT/FORS2 spectrum around the Ly$\alpha$ absorption line.
Green, magenta, blue, and red colors correspond to absorption lines of
$z=1.043$, $1.952$, $1.962$, and $1.986$ systems, respectively.  The
damped Ly$\alpha$ absorption of VLT spectrum is better matched to two
absorption systems of ${\rm log}\,N({\rm HI})=20.5$ at $z_3$ and $z_4$
(blue line) than a single absorption profile with $z=1.975$ and ${\rm
log}\,N({\rm HI})=21.2$ (red line).
\label{fig:spec}}
\end{figure}

In the first spectrum, shown in Figure~\ref{fig:spec}, we identify
four absorption systems at $z_1=1.04329\pm 0.00001$, $z_2=1.95260\pm
0.00004$, $z_3=1.96337\pm 0.00014$, and $z_4=1.98691\pm 0.00015$.  The
absorbers $z_1$, $z_3$ and $z_4$ were identified in \citet{kgr+04}.
We identify the additional absorber $z_2$ from its \ion{C}{4} lines.
Finally , the uncertain lines at 5494 and 5517 \AA\ noted in the VLT
spectrum are revealed as \ion{Al}{3} lines of $z_3$ here.

We further compared the equivalent widths and redshifts of all lines
detected in our two ESI spectra to search for time-dependent effects.
Within the uncertainties of about $10\%$ in equivalent width and about
10 km s$^{-1}$ in redshift, we find no systematic changes for any of
the low- and high-ionization lines.

We measured the column densities of the various absorption lines using
the apparent optical depth (AOD) method \citep{ss91}: 
\begin{equation}
\tau_{a}(v)=-{\rm ln}[I_{\rm obs}(v)/I_{c}(v)],
\end{equation}
where $I_{\rm obs}$ and $I_{c}$ are the observed and continuum
intensities, respectively.  The apparent column density is given by:
\begin{equation}
N_{a}(v)=3.768\times 10^{14}f \lambda\tau_{a}(v)\,\,{\rm cm^{-2}(km ~ 
s^{-1})^{-1}},
\end{equation}
where $f_\lambda$ and $\lambda$ are the oscillator strengths and
wavelengths of the absorption lines, respectively.  The total apparent
column density is given by $N_{a}=\int^{v_u}_{v_l}N_{a}(v) dv$ between
velocity limits $v_l$ and $v_u$.  The measured rest-frame equivalent
widths and inferred column densities for each line are given in
Tables~\ref{tab:z1z2} and \ref{tab:z3z4}, and the final adopted 
columns are summarized in Table~\ref{tab:columns}.

\section{Results}
\label{sec:res}

\begin{figure}
\epsscale{1}
\plotone{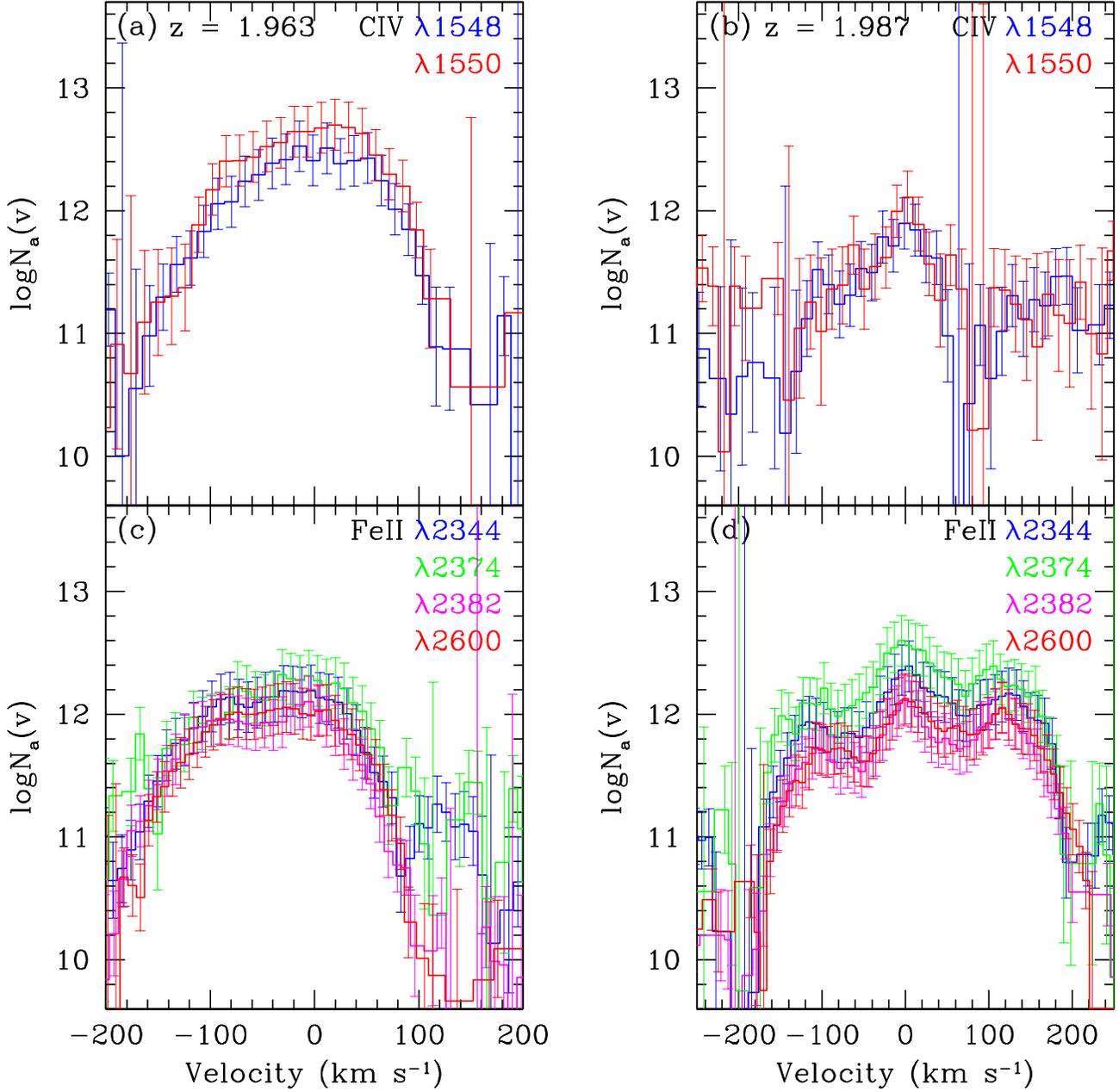}
\caption{Apparent column density, $N_{a}(v)$, plots for $z_3$ (panels
a and c) and $z_4$ (panels b and d).  High-ionization lines such as
\ion{C}{4} show a single asymmetric peak, while low-ionization lines
such as \ion{Fe}{2} exhibit multiple sub-components.
\label{fig:column}}
\end{figure}

The absorption systems $z_3$ and $z_4$ are separated by $2360$ km
s$^{-1}$ and exhibit strong low-ionization lines of \ion{Fe}{2},
\ion{Si}{2}, \ion{Al}{2}, and \ion{Mg}{2}, as well as high-ionization
lines of \ion{Si}{4}, \ion{C}{4}, and in the case of $z_3$
\ion{Al}{3}.  The kinematic structure of these lines is apparent in
the column density profiles shown in Figure~\ref{fig:column}.  The
low-ionization lines show two and three sub-components spanning a
total velocity range of about $300$ and $400$ km s$^{-1}$ for $z_3$
and $z_4$, respectively.  This is similar to the absorption profiles
found in previous echelle spectra of, for example, GRB\,051111
\citep{pbf+06,pcb06}.  On the other hand, the high-ionization lines
show only a single asymmetric peak, which coincides in velocity with
the strongest sub-component of the low-ionization lines
(Figure~\ref{fig:column}).

\subsection{Abundances}
\label{sec:abund}

\begin{figure}
\epsscale{1} 
\plotone{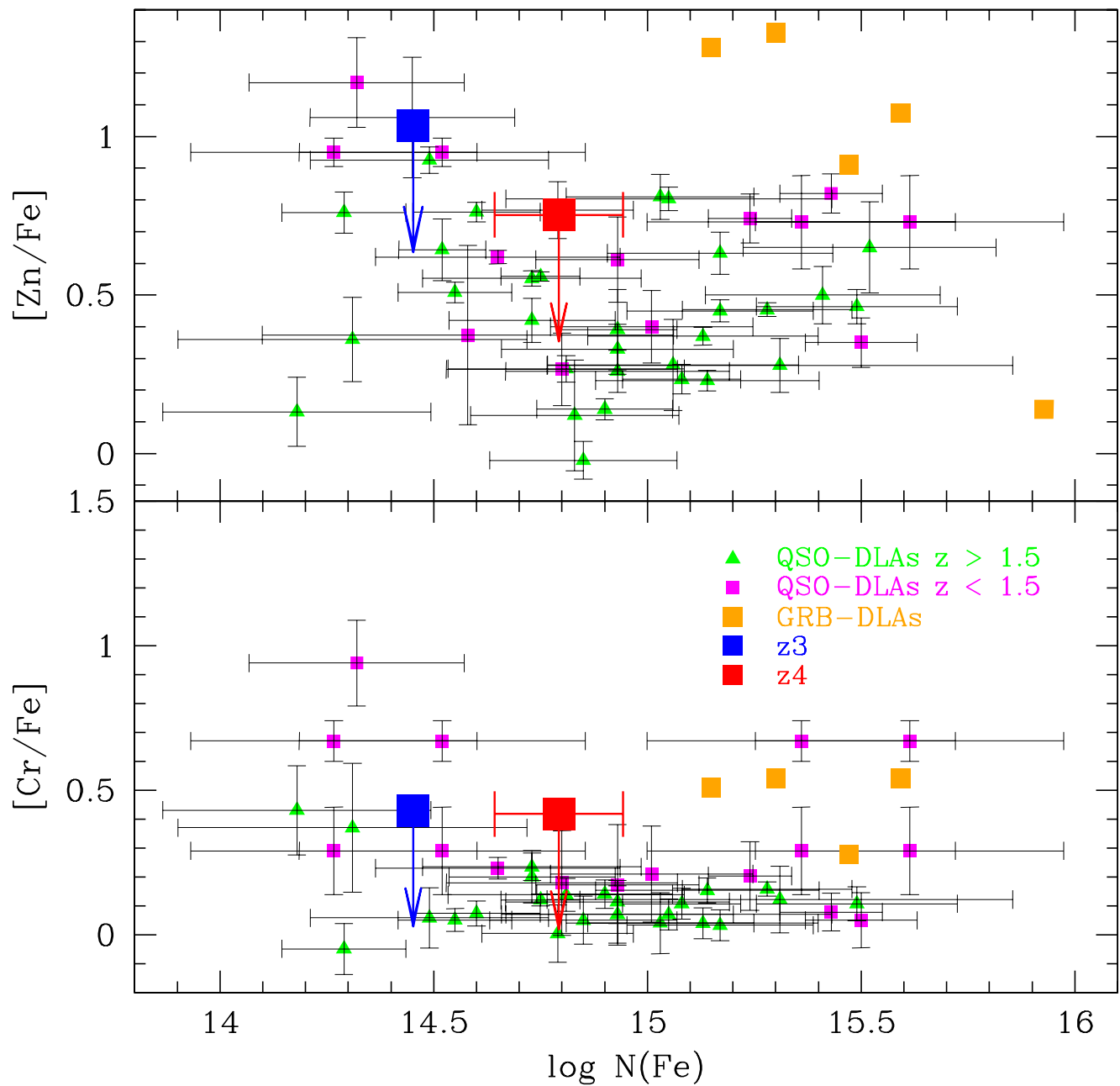}
\caption{{\it Top}): [Zn/Fe] vs.~${\rm log}\,N({\rm FeII})$ for $z_3$
and $z_4$, as well as previously-observed GRB-DLAs (orange squares)
and QSO-DLAs (magenta for $z<1.5$ and green for $z>1.5$) from
\citet{mkk+06}.  ({\it Bottom}): Same but for [Cr/Fe].  Note that
the error bars for the QSO-DLAs are over-estimated since the values
were transformed with the assumption of non-correlated errors.
\label{fig:fe}}
\end{figure}

Our ability to measure the chemical abundances of the $z_3$ and $z_4$
absorbers is somewhat limited by line saturation and the uncertainty
in the hydrogen column density.  Still, for the low-ionization species
we are able to derive robust column densities for \ion{Fe}{2}, as well
as limits for \ion{Cr}{2}, \ion{Zn}{2}, and excited fine-structure
states of \ion{Si}{2} and \ion{Fe}{2}.  Compared to most previous GRB
absorption systems (e.g., \citealt{sf04}), the column densities of
\ion{Fe}{2} are at least 0.5 dex lower, ${\rm log}\,N({\rm
FeII})=14.45$ and $14.8$, respectively (see Figure~\ref{fig:fe}).
Based on the non-detection of \ion{Fe}{2}$\lambda 2260$, we find ${\rm
log}\,N ({\rm FeII})<14.9$ and $<15.1$, respectively, indicating that
saturation is not a significant problem.

The limits on \ion{Cr}{2} and \ion{Zn}{2} are derived by applying the
AOD method to the appropriate velocity range around
\ion{Cr}{2}$\lambda 2056$ and \ion{Zn}{2}$\lambda 2026$.  We find
${\rm log}\,N({\rm ZnII})<12.6$ and $<12.7$, and ${\rm log}\,N({\rm
CrII})<13.1$ and $<13.4$ for $z_3$ and $z_4$, respectively.  Thus,
${\rm [Cr/Fe]}<0.4$ for both systems, in good agreement with the
values measured for QSO absorbers of $\sim 0.1-0.3$
(Figure~\ref{fig:fe}).  Similarly, ${\rm [Zn/Fe]}<1.0$ and $<0.75$ for
$z_3$ and $z_4$, respectively, placing it within the range of values
found for QSO absorbers, ${\rm [Zn/Fe]}\sim 0-1$ (e.g.,
\citealt{mkk+06}).  The majority of previous GRB absorbers have ${\rm
[Zn/Fe]}\gtrsim 1.0$, which has been interpreted as an indication for
dust depletion\footnotemark\footnotetext{But see \citet{pro06} for the
impact of spectral resolution on these conclusions.} \citep{sf04}
based on the fact that zinc is a non-refractory element, while iron
can be strongly depleted in dust grains.  In the case of \grb,
multi-band afterglow observations and polarization measurements
\citep{kgr+04} point to a low dust-to-gas ratio along the line of
sight.  This suggests that our values of [Zn/Fe] and [Cr/Fe] are
meaningful upper limits.

To derive the metallicity of $z_3$ and $z_4$ we re-analyzed the
VLT/FORS2 spectrum presented in \citet{kgr+04} following the
procedures outlined in \S\ref{sec:obs}.  The signal-to-noise at the
location of the Ly$\alpha$ absorption feature is low, $\sim 3$ per
resolution element, but we find that the line profile is best fit with
a pair of absorbers at the redshifts of $z_3$ and $z_4$, each with
${\rm log}\,N$(\ion{H}{1})$\sim 20.5$ and an uncertainty of about 0.3
dex; a more marginal fit for a single absorber has $z=1.975$ and ${\rm
log}\,N$(\ion{H}{1})$\sim 21.2$ (see inset in Figure~\ref{fig:spec}).
Thus, using the limit on \ion{Zn}{2}, we find a metallicity ${\rm
[Zn/H]}\lesssim -0.5$ and $\lesssim -0.4$ for $z_3$ and $z_4$,
respectively.  Assuming iron is weakly depleted as discussed above we
find ${\rm [Fe/H]}\sim -1.5$ and $\sim -1.2$, respectively.  These
values are at the low end of the distribution for GRB-DLAs
\citep{bpc+06}.

\subsection{Kinematics}

Our primary interest is in understanding the nature of the $z_3$ and
$z_4$ systems in light of their 2400 km s$^{-1}$ velocity separation.
This is similar to the multiple absorption features spanning $\sim
3000$ km s$^{-1}$ observed in the spectrum of GRB\,021004, which have
been interpreted to arise in the circumstellar wind of the massive
progenitor star \citep{mfh+02,sgh+03,mhc+03,fdl+05,swh+05}.  An
alternative explanation is that $z_3$ could be an intervening system,
while the absorption lines of $z_4$ arise from interstellar gas in the
GRB host galaxy.  Such multiple intervening absorption systems have
been found in QSO spectra \citep{el01,le03,mcd+05}.

Several lines of evidence suggest that the circumstellar wind
scenario\footnotemark\footnotetext{Or for that matter other phenomena
local to the burst such as a superbubble produced by stellar winds and
supernovae in the burst's star forming region with a scale of a few pc
(e.g., \citealt{mhc+03}).} is an unlikely explanation.  First, unlike
in the case of GRB\,021004 we do not detect an intermediate velocity
(few hundred km s$^{-1}$) absorber.  Such an absorber is expected to
arise from the interaction of the WR wind with the red supergiant wind
\citep{vlg05}.  Still, this may not be a significant problem for the
circumstellar model if the GRB explosion occurred more than $\sim
10^5$ yr after the onset of the WR phase, since on this timescale the
shell dissipates into the circumstellar bubble \citep{vlg05}.  If this
was in fact the case, then it would appear that the progenitor of
GRB\,021004 exploded earlier in its evolution compared to the
progenitor of \grb.

However, a more difficult problem to overcome for the circumstellar
model is the presence of strong low-ionization lines in spite of the
progenitor and the burst ionizing radiation.  This problem has already
been noted for GRB\,021004, in particular the detection of outflowing
hydrogen gas kinematically coincident with highly ionized gas
(\ion{C}{4}, \ion{Si}{4}).  From the observed brightness of the burst,
$V\approx 19$ mag at $t=4.6$ hr \citep{kgr+04}, we expect species such
as \ion{Mg}{2} (with $\sigma_{\rm ph}=7\times 10^{-20}$ cm$^{2}$ at
$E=15$ eV) to be ionized out to a distance of about 50 pc from the
burst (following the formulation of \citealt{pcb06}) in the absence of
shielding by a large column of neutral hydrogen gas.  The fast WR wind
is expected to span only $\lesssim 10$ pc, well within the expected
radius of the Str\"omgren sphere, $\sim 20-30$ pc \citep{vlg05}, and
therefore well inside of the region of $\sim 50$ pc in which
\ion{Mg}{2} and \ion{Fe}{2} will be ionized in the absence of neutral
hydrogen.

In the case of GRB\,021004, the ratio of \ion{Fe}{2} to \ion{Si}{4}
was ${\rm log}[N({\rm FeII})/N({\rm SiIV})]\approx -1.9$ and $-1.2$
for the zero-velocity and the $-2800$ km s$^{-1}$ systems,
respectively \citep{fdl+05}.  However, here we find the opposite
situation, namely \ion{Fe}{2} is over-abundant relative to
\ion{Si}{4}, with values of $\sim 0.1$ and $1.4$ for $z_3$ and $z_4$,
respectively.

Even if neutral hydrogen could survive near the progenitor star, the
measured column density of ${\rm log}\,N({\rm HI})\sim 20.5$ cm$^{-2}$
(\S\ref{sec:abund}) is more than four orders of magnitude larger than
that of GRB\,021004, and requires the presence of $\sim 30$ M$_\odot$
of hydrogen even if we limit the size of the WR wind to $\sim 1$ pc.
There is no clear scenario that can allow for such a large mass of
hydrogen within the WR wind.  The inferred column density of
\ion{Fe}{2} is equally high in this context.  For a mass loss rate of
$\sim 10^{-5}$ M$_\odot$ yr $^{-1}$ and an iron mass fraction of about
$10^{-3}$ \citep{gh05} the expected column density is only about
$2\times 10^{12}$ cm$^{-2}$ if the iron is distributed uniformly over
a radius of 1 pc.  Clumping may lead to regions of significantly
higher column density, but in this case we expect only partial
covering, as opposed to the observed zero transimission at the bottom
of the absorption features.

Finally, the absence of absorption features from fine-structure
\ion{Fe}{2} and \ion{Si}{2} suggests that the gas is located far from
the burst, in the framework of radiative pumping, and/or has a
relatively low density if we consider collisional excitation.  We find
a limit of ${\rm log}\,N({\rm FeII^*})<12.7$ and $<12.6$ for $z_3$ and
$z_4$, respectively, from the non-detection of \ion{Fe}{2}$^*\lambda
2396$.  This corresponds to a ratio of $N({\rm FeII^*})/N({\rm
FeII})<10^{-1.8}$ and $<10^{-2.2}$ for the two absorbers.  In
comparison, for GRB\,051111 the corresponding value was about
$10^{-1.3}$ \citep{pbf+06,pcb06}.  For $z_3$ the limit on
fine-structure \ion{Si}{2}$^*$ is $N({\rm SiII^*})/N({\rm SiII})
<10^{-1.3}$, somewhat lower than the detected ratio of $10^{-1.2}$ in
the case of GRB\,051111 \citep{pbf+06,pcb06}; \ion{Si}{2}$^*$ for
$z_4$ is blended with the \ion{C}{4}$\lambda 1550$ line of $z_2$,
preventing a useful limit on the column density.

In the context of collisional excitation, the limit on $N({\rm
SiII^*})/N({\rm SiII})$, indicates a hydrogen volume density of
$\lesssim 5\times 10^3$ cm$^{-3}$ for $T=10^3$ K and an electron
fraction of $<10^{-4}$ \citep{sv02}.  In the case of radiative
pumping, the non-detection of fine-structure absorption suggests that
the gas is located $\gtrsim 100$ pc away from the burst \citep{pcb06},
much beyond the fast WR wind.  This limit is further supported by the
lack of changes in line equivalent widths (\S\ref{sec:obs}).

We note that in the case of GRB\,021004 it has been argued that it may
be possible to maintain low-ionization species (and hydrogen) in the
circumstellar environment if the GRB jet was for example structured
\citep{swh+05}, or as a result of shielding and mixing \citep{mhc+03}.
However, these explanations are at best speculative for the case of
GRB\,021004, and are unlikely to work in the more extreme case of high
column density low-ionization species in \grb.

Based on these various lines of reasoning, we conclude that the
blue-shifted system does not arise in the progenitor wind. Instead we conclude 
that $z_3$ and $z_4$ are multiple DLAs, remarkably
similar to the multiple intervening absorbers found by
\citet{el01,le03} at a similar redshift and with a similar velocity of
separation of $\sim 2000$ km s$^{-1}$.  In particular, the $z_3$
absorber is similar to absorbers CTQ247C observed by \citet{le03} and
DLA B introduced in \citet{el01}. Both of these have a similar
hydrogen column density with no detectable \ion{Cr}{2} and \ion{Zn}{2}
lines.

\section{Discussion}
\label{sec:disc}

We find four absorption systems in the spectrum of \grb, of which the
two highest redshift systems are separated by only 2400 km s$^{-1}$.
Absorber $z_2$ is separated by about 3400 km s$^{-1}$ from $z_4$ and
1100 km s$^{-1}$ from $z_3$, but is detected only in \ion{C}{4}.  As
such it most likely arises in the halo of an intervening galaxy.
Although $z_3$ was previously suspected to arise in the outflowing
circumstellar gas surrounding the progenitor star \citep{kgr+04}, we
conclude that this scenario is unlikely based on the presence of high
column density low-ionization species, the lack of excited
fine-structure lines, the large column density of iron, and the
possible presence of large column density of neutral hydrogen.  A more
likely scenario is that system $z_3$ is an intervening absorber.

This conclusion leaves only GRBs 021004 and 050505 as showing
potential evidence for wind outflows from the progenitor star.  In the
former case, the presence of high velocity neutral hydrogen, as well
as \ion{Mg}{2} and \ion{Fe}{2}, presents a challenge in the context of
WR wind models and the ionizing radiation of the burst \citep{swh+05}.
In the latter case, only high-ionization \ion{C}{4} is detected in
outflow \citep{bpc+06}, removing the problem of explaining the
presence of low-ionization lines, but the low resolution of the
spectrum prevents a detailed study.  Given the influence of the
kinematic and ionization structure of the wind in the pre-explosion
environment, as well as the effects of the GRB radiation field itself,
it is of prime importance to make model predictions for the type and
column density of various ionic species that can be observed in the
afterglow spectrum.  Such models can be used to assess whether any of
the outflowing systems observed to date are in fact signatures of the
stellar wind.

Assessing $z_3$ as an intervening absorber, we note that both $z_3$
and the host galaxy absorber $z_4$ exhibit similar abundances and
kinematics compared to previous GRB-DLAs and multiple DLAs found in
QSO sight lines.  Assuming that iron is weakly depleted (as suggested
by several lines of evidence, see \S\ref{sec:abund}) the gas
metallicity is somewhat lower than in most previous GRB-DLAs, ${\rm
[Fe/H]}\sim -1.5$ and $\sim -1.2$, respectively.

Since $z_3$ most likely represents an intervening galaxy, we may be
able to identify it in deep imaging around the GRB position, and
spectroscopy of possible intervening galaxies in the field.  Previous
narrow-band imaging observations aimed at finding Ly$\alpha$ emission
in the range $z\approx 1.94-2.0$, have uncovered three galaxies that
may have similar redshifts to the $z_2$ and $z_3$ absorbers
\citep{jbf+05}.  No source was detected at the GRB position.  These
putative counterparts are faint and it is reasonable to interpret both
the undetected GRB host galaxy and the intervening systems as low mass
galaxies with low metallicity and dust content.

The remaining open question is whether there is a link between the GRB
host absorber and the intervening systems.  It has recently been shown
that the probability of finding multiple intervening absorbers toward
GRBs is larger compared to QSO sight lines \citep{ppc+06}.  However,
the majority of these systems are separated by $>10^4$ km s$^{-1}$ and
are unlikely to be physically connected.  With a velocity separation
of 2400 km s$^{-1}$ it is conceivable that $z_3$ and the GRB host are
part of the same large-scale structure, as has been suggested for the
multiple DLAs in QSO sight lines \citep{le03}.  Future observations of
the environment of \grb, as well as similar absorption systems may
shed light on the dependence of the number and velocity difference of
intervening systems on the properties of GRB host galaxies and their
large-scale environments.

\acknowledgments
We thank J.~Stone, and B.~Draine for helpful discussions, and
J.~Meiring and collaborators for providing us with the data used in
Figure 3.  M.-S.~S.~acknowledges support from the Observatories of the
Carnegie Institution of Washington. 
E.B.~is supported by NASA through Hubble
Fellowship grant HST-01171.01 awarded by STSCI, which is operated by
AURA, Inc., for NASA under contract NAS5-26555. P.C. acknowledges
support provided by NASA LTSA grant NAG5-11714. The data presented herein were obtained 
at the W.M. Keck Observatory, which is 
operated as a scientific partnership among the California Institute of Technology, 
the University of California and the National Aeronautics and Space Administration. 
The Observatory was made possible by the generous financial support of the W.M. Keck Foundation.
The authors wish to 
recognize and acknowledge the very significant cultural role and 
reverence that the summit of Mauna Kea has always had within the 
indigenous Hawaiian community.  We are most fortunate to have the 
opportunity to conduct observations from this mountain.

\clearpage
\begin{deluxetable}{lllll}
\tablecolumns{5}
\tabcolsep0.2in\footnotesize
\tablewidth{0pc}
\tablecaption{Line Identification for $z_1$ and $z_2$
\label{tab:z1z2}}
\tablehead {
\colhead {$\lambda_{\rm obs}$}         &
\colhead {Line}                        &
\colhead {$f_{ij}$}                    &
\colhead {$W_0$}                       &
\colhead {${\rm log}\,N$}              \\
\colhead {(\AA)}                       &
\colhead {}                            &
\colhead {}                            &
\colhead {(\AA)}                       &
\colhead {(cm$^{-2}$)}
}
\startdata
\multicolumn{5}{c}{$z_1=1.04329$} \\
\hline
5713.77 & \ion{Mg}{2} 2796.35 & 0.6123 & $0.30\pm 0.01$ & $12.95\pm 0.02$ \\
5728.38 & \ion{Mg}{2} 2803.53 & 0.3054 & $0.27\pm 0.01$ & $13.16\pm 0.02$ \\
\hline
\hline
\multicolumn{5}{c}{$z_2=1.95260$} \\
\hline
4571.11 & \ion{C}{4} 1548.20 & 0.1908 & $0.12\pm 0.01$ & $13.54\pm 0.04$ \\
4578.85 & \ion{C}{4} 1550.77 & 0.0952 & $0.09\pm 0.01$ & $13.77\pm 0.06$ 
\enddata
\tablecomments{Absorption features of the intervening redshift 
systems $z_1$ and $z_2$ identified in our spectrum of \grb.
Uncertainties are $1\sigma$.}
\end{deluxetable}

\clearpage
\begin{deluxetable}{llllllll}
\tablecolumns{8}
\tabcolsep0.05in\footnotesize
\tablewidth{0pc}
\tablecaption{Line Identification for $z_3$ and $z_4$
\label{tab:z3z4}}
\tablehead {
\colhead {}                            &
\colhead {}                            &
\multicolumn{3}{c}{$z_3=1.96337$}      &
\multicolumn{3}{c}{$z_4=1.98691$}      \\\cline{3-5}\cline{6-8}
\colhead {Line}                        &
\colhead {$f_{ij}$}                    &
\colhead {$\lambda_{\rm obs}$}         &
\colhead {$W_0$}                       &
\colhead {${\rm log}\,N$}              &
\colhead {$\lambda_{\rm obs}$}         &
\colhead {$W_0$}                       &
\colhead {${\rm log}\,N$}              \\
\colhead {}                            &
\colhead {}                            &
\colhead {(\AA)}                       &
\colhead {(\AA)}                       &
\colhead {(cm$^{-2}$)}                 &
\colhead {(\AA)}                       &
\colhead {(\AA)}                       &
\colhead {(cm$^{-2}$)} 
}
\startdata
\ion{Si}{4} 1393.75 & 0.5280 & 4130.03 & $0.751\pm 0.008$ & $14.28\pm 0.02$ & 4163.70 & $0.190\pm 0.020$ & $13.44\pm 0.05$ \\
\ion{Si}{4} 1402.77 & 0.2620 & 4156.77 & $0.681\pm 0.009$ & $14.41\pm 0.02$ & \nod    & \nod             & \nod            \\
\ion{Si}{2} 1526.71 & 0.1270 & 4523.30 & $0.784\pm 0.007$ & $14.74\pm 0.02$ & 4558.77 & $0.969\pm 0.012$ & $14.75\pm 0.01$ \\
\ion{C}{4}  1548.20 & 0.1908 & 4587.62 & $0.848\pm 0.008$ & $14.64\pm 0.02$ & 4625.02 & $0.318\pm 0.018$ & $13.97\pm 0.03$ \\
\ion{C}{4}  1550.77 & 0.0952 & 4595.50 & $0.755\pm 0.009$ & $14.83\pm 0.02$ & 4632.56 & $0.199\pm 0.018$ & $14.06\pm 0.04$ \\
\ion{Fe}{2} 1608.45 & 0.0580 & 4764.75 & $0.305\pm 0.007$ & $14.45\pm 0.02$ & 4802.22 & $0.637\pm 0.008$ & $14.81\pm 0.02$ \\
\ion{Al}{2} 1670.79 & 1.8800 & 4951.64 & $0.811\pm 0.007$ & $13.51\pm 0.02$ & 4989.84 & $0.904\pm 0.011$ & $13.45\pm 0.01$ \\
\ion{Al}{3} 1854.72 & 0.5390 & 5495.90 & $0.469\pm 0.006$ & $13.57\pm 0.02$ & \nod    & \nod             & \nod            \\
\ion{Al}{3} 1862.79 & 0.2680 & 5519.88 & $0.291\pm 0.006$ & $13.62\pm 0.02$ & \nod    & \nod             & \nod            \\
\ion{Fe}{2} 2344.21 & 0.1140 & 6946.88 & $0.887\pm 0.004$ & $14.36\pm 0.01$ & 7001.30 & $1.399\pm 0.005$ & $14.58\pm 0.01$ \\
\ion{Fe}{2} 2374.46 & 0.0313 & 7036.33 & $0.378\pm 0.006$ & $14.45\pm 0.02$ & 7091.71 & $0.788\pm 0.008$ & $14.79\pm 0.01$ \\
\ion{Fe}{2} 2382.77 & 0.3200 & 7061.33 & $1.320\pm 0.003$ & $14.22\pm 0.01$ & 7116.45 & $1.828\pm 0.005$ & $14.32\pm 0.01$ \\
\ion{Fe}{2} 2586.65 & 0.0691 & \nod    & \nod             & \nod            & 7725.35 & $1.367\pm 0.006$ & $14.67\pm 0.01$ \\
\ion{Fe}{2} 2600.17 & 0.2390 & 7705.29 & $1.335\pm 0.003$ & $14.24\pm 0.01$ & 7765.60 & $1.929\pm 0.007$ & $14.37\pm 0.01$ \\
\ion{Mg}{2} 2796.35 & 0.6123 & 8286.77 & $1.939\pm 0.006$ & $14.12\pm 0.02$ & 8351.53 & $2.647\pm 0.015$ & $14.17\pm 0.01$ \\
\ion{Mg}{2} 2803.53 & 0.3054 & 8307.97 & $1.778\pm 0.004$ & $14.42\pm 0.01$ & 8373.01 & $2.478\pm 0.006$ & $14.39\pm 0.01$
\enddata
\tablecomments{Absorption features of the redshift systems $z_3$ 
and $z_4$ identified in our spectrum of \grb.  Uncertainties are 
$1\sigma$.}
\end{deluxetable}

\clearpage
\begin{deluxetable}{lcc}
\tablecolumns{3}
\tabcolsep0.2in\footnotesize
\tablewidth{0pc}
\tablecaption{Ionic Column densities for $z_3$ and $z_4$
\label{tab:columns}}
\tablehead {
\colhead {}                            &
\colhead {$z_3$}                       & 
\colhead {$z_4$}                       \\\cline{2-3}
\colhead {Ion}                         &
\colhead {${\rm log}\,N$}              &
\colhead {${\rm log}\,N$}              
}
\startdata
\ion{C}{4}      & $>14.83$ & $14.06$       \\
\ion{Mg}{2}     & $>14.42$ & $>14.39$      \\
\ion{Al}{2}     & $13.51$  & $13.45$       \\
\ion{Al}{3}     & $13.62$  & \nod          \\
\ion{Si}{2}     & $>14.74$ & $>14.75$      \\
\ion{Si}{2}$^*$ & $>14.74$ & $>14.75$      \\
\ion{Si}{4}     & $>14.41$ & $13.44$       \\
\ion{Cr}{2}     & $<13.1$  & $<13.4$       \\
\ion{Fe}{2}     & $14.45$  & $14.79-15.13$ \\
\ion{Fe}{2}$^*$ & $<12.7$  & $<12.6$       \\
\ion{Zn}{2}     & $<12.6$  & $<12.7$ 
\enddata
\tablecomments{Ionic column densities of the $z_3$ and $z_4$ 
absorption systems.}
\end{deluxetable}

\clearpage

\clearpage

\clearpage

\end{document}